\documentclass[aps,prl,twocolumn,groupedaddress,showpacs]{revtex4-1}
\pdfoutput=1

\usepackage{amsmath}
\usepackage{graphicx}
\usepackage{bm}

\usepackage{bbold}

\usepackage{multirow}

\usepackage[extra]{tipa}

\newcommand{\ma}{\mathbb{1}}
\newcommand{\mb}{\mathbb{2}}
\newcommand{\mc}{\mathbb{3}}
\newcommand{\md}{\mathbb{4}}

\newcommand{\la}{\langle}
\newcommand{\ra}{\rangle}
\newcommand{\ua}{\uparrow}
\newcommand{\da}{\downarrow}

\newcommand{\vk}{{\bf k}}

\newcommand{\vx}{{\bf x}}
\newcommand{\vd}{\bm{\delta}}

\usepackage{color}
\usepackage{color}

\newcommand{\io}{i\omega_n}

\newcommand{\e}{\epsilon}

\usepackage{hyperref}
\hypersetup{%
  breaklinks = {true},
  citecolor = {blue},
  colorlinks = {true},
  linkcolor = {red},
  pdfauthor = {\textcopyright\ Johan Nilsson},
  pdfcreator = {\LaTeX\ and \flqq hyperref\frqq},
  pdffitwindow = {true},
  pdfmenubar = {true},
  pdfpagelayout = {SinglePage},
  pdfstartview = {Fit},
  pdftoolbar = {true},
  plainpages = {false},
}

\begin{document}

\title{Free fermion description of a paramagnetic Mott insulator}

\author{Johan Nilsson}
%\author{}
\affiliation{Department of Physics, University of Gothenburg, 412 96 Gothenburg,  Sweden}
\author{Matteo Bazzanella}
\affiliation{Department of Physics, University of Gothenburg, 412 96 Gothenburg,  Sweden}

\date{July 2014}

\begin{abstract}

A scheme is presented that enables a description of a paramagnetic Mott insulator in terms of free fermions. The main idea is to view the physical fermions as a part of a multi-band system and to allow for a correlation between the physical fermions and the auxiliary ones. Technically this is implemented through a non-linear canonical transformation, which is conveniently formulated in terms of Majorana fermions. The transformed Hamiltonian is in the next stage approximated with a free fermion theory. The approximation step is variational and provides an upper bound on the ground state energy at zero or the Free energy at finite temperature.

\end{abstract}

\maketitle

{\it Introduction.} To understand and describe the correlation driven Mott metal-insulator transition is a widely known difficult problem \cite{RMP_MIT1998,Gebhard_review}. Identifying and writing down a simple theory for the unstable paramagnetic Mott insulator (PMI) fixed point would go a long way towards a qualitative solution of this problem \cite{Phillips_RMP}. Indeed, if this fixed point theory was known, it should be possible to study how different perturbations take the system away from it. For example, the PMI is typically expected to be unstable towards long-range ordered antiferromagnetism in dimensions {$d \geq 3$ for low temperatures at half filling. Going instead away from half filling superconductivity should appear at some point if the cuprates are to be described as doped Mott insulators \cite{Lee_Nagaosa_Wen_RMP}.

The fixed point theories for metals and band insulators are free fermion theories and therefore both simple to understand and calculate with. Many theories for the PMI have been proposed over the years.
The Gutzwiller wave function \cite{Gutzwiller1965,BrinkmanRice1970}, for example, provides a simple physically motivated variational wave function for the PMI. A recent body of works describes the Mott insulator in terms of fermions and charged bosons \cite{Phillips_RMP}. Approximate theories can also be obtained by Green's function decoupling schemes of different kinds, see e.g. Ref.~\cite{Gebhard_review}.
Unfortunately none of these descriptions are in terms of free fermions only. Indeed, a Mott insulator is by definition insulating, although according to simple band theory it should be a metal, so it is often claimed that the fixed point theory of a PMI can not be described in terms of free fermions.
%, see for example \cite{Phillips_RMP}. 
In this work, however, we propose such a free fermion description.

We consider the Mott insulator in the framework of the single band Hubbard model \cite{HubbardI}. The ground state of the half-filled Hubbard model in $d=1$ is a paramagnetic Mott insulator for any value of the repulsion. This can be understood both from the exact solution \cite{LiebWu1968} and bosonization arguments \cite{Giamarchi_Book}. In $d \rightarrow \infty$ the Hubbard model can be solved exactly with dynamical mean field theory (DMFT) \cite{DMFT_RMP1996}, which gives a metal-insulator transition. In this case one can get a description of a PMI by enforcing paramagnetism by hand or by introducing frustration. The main feature of DMFT is that the effect of the interaction is described by a local frequency-dependent self-energy $\Sigma(\omega)$. The single particle Green's function of DMFT could be obtained by considering the physical fermions to be a part of a multi-band system. Integrating out the auxiliary bands (note that this fermion bath is different from that used to parametrize the ``Weiss function'' of the corresponding impurity problem \cite{DMFT_RMP1996}) one obtains the local $\Sigma(\omega)$ of DMFT if the additional bands only have local dynamics. The key feature of DMFT is that it provides a self-consistent scheme to determine the parameters of the auxiliary multi-band system. An alternative, conceptually very attractive scheme, that relies on a dynamical variational principle for the self-energy, has also been proposed \cite{Potthoff2003}. In principle the multi-band system should have an infinite number of bands to be able to describe any $\Sigma(\omega)$, but keeping a few bath sites often gives qualitatively good results.
This way of looking at DMFT, which is not how it is conventionally presented, allows one to make connections with other popular methods such as the density matrix renormalization group (DMRG) and holography. In these techniques the physical density matrix or boundary degrees of freedom are obtained by integrating out the unphysical auxiliary degrees of freedom: i.e., taking the trace over the additional matrix space in the matrix product state representation of the DMRG fixed point \cite{Ostlund_Rommer1995,RMP_DMRG2005}, or getting rid of the bulk degrees of freedom in holography.

In this letter we propose an alternative way of introducing auxiliary degrees of freedom for fermion theories. Like in the viewpoint on DMFT mentioned above, we consider the physical fermions to be a part of a larger $M$-band system. Note that this is opposite to the more familiar approach of reducing a physical interacting $M$-band system down to a single-band Hubbard model \cite{ZhangRice1988}. We introduce no dynamics for the auxiliary system to start with, although it is certainly allowed to do so and this would provide additional variational freedom. The crucial step is to allow for a correlation between the physical and auxiliary systems without any explicit coupling in the Hamiltonian. Technically this is implemented by a direct product of local non-linear canonical transformations on the fermions that describe the system \cite{OstlundMele1991,Ostlund2007,StellanHansAnders2005}. Once this is done we determine the best free fermion theory to describe the system with. Here ``best'' is with respect to the variational upper bound on the ground state energy at $T=0$ or the Free energy at $T > 0$. In this way we are able to describe a PMI as a band insulator in the extended multi-band system.

{\it Hubbard model.}
Our method is applicable to generic lattice fermion models, but for definiteness we focus on the single-band Hubbard model on simple cubic lattices in $d$ dimensions at half filling, parametrized by the nearest neighbor hopping energy $t > 0$ and the local interaction strength $U > 0$.
The Hamiltonian is
\begin{equation}
H = -t \sum_{\la i,j \ra,\sigma}
c_{i 1 \sigma}^\dagger c_{j 1 \sigma}^{\,}
+U \sum_{i}  
%\bigl( n_{i1\ua} - \frac{1}{2} \bigr) 
%\bigl(n_{i1\da} - \frac{1}{2} \bigr),
\Bigl( n_{i1\ua} - \frac{1}{2} \Bigr) 
\Bigl(n_{i1\da} - \frac{1}{2} \Bigr),
%c_{i 1\ua}^{\dagger} c_{i 1\ua}^{\,}
%c_{i 1\da}^{\dagger} c_{i 1\da}^{\,},
\label{eq:HubbardHamiltonian}
\end{equation}
where, as is conventional, the first sum goes over nearest neighbors and $n_{i1\sigma} = c_{i 1\sigma}^{\dagger} c_{i 1\sigma}^{\,}$.
The main idea behind our method is to view this Hamiltonian as a part of a multi-band system, where the rest of the bands have no dynamics to start with. This is already put into the notation we use for the fermion creation operators $c_{i \mu \sigma}^\dagger$: $i = 1,\ldots,N$ is the spatial unit cell index, $\mu=1,\ldots ,M$ the band index, and $\sigma=\ua,\da$ the spin index.

%The next step is to perform a non-linear canonical transformation that allows for a mixing of the different bands \cite{Ostlund2007,nilsson2011a,Bazzanella2014b}. Once this is done we will approximate the resulting theory with a free fermion theory in terms of the transformed fermions. This is expected to be good if it is possible to describe the most important terms in the Hamiltonian in terms of free fermions.

{\it Majorana fermion representation.} To perform and classify the non-linear transformations that we are going to use, it is convenient (but not necessary) to first reformulate the theory in terms of Majorana fermions \cite{Bazzanella2014b}. Let us define Majorana operators $\gamma_{i \mu a}$ ($a=\ma, \mb, \mc, \md$ is the Majorana flavor index) via
\begin{align}
\begin{split}
c^\dagger_{j \mu \ua} &= e^{i \chi_{j\mu}} \frac{\gamma_{j \mu \ma} + i \gamma_{j \mu \mb}}{2} ,
\\
c^\dagger_{j \mu \da} &= e^{i \chi_{j\mu}} \frac{\gamma_{j \mu \mc} + i \gamma_{j \mu \md}}{2} .
\end{split}
\label{eq:majoranadef}
\end{align}
We use the usual definition of Majoranas in condensed matter physics, i.e., they are real $\gamma^{\dagger}_{i \mu a}=\gamma^{}_{i \mu a}$ and satisfy  the Clifford algebra
$\{ \gamma_{i \mu a}, \gamma_{j\nu b} \}
= 2 \delta_{ij} \delta_{\mu\nu}\delta_{ab}$.
Within each unit cell we use the phase convention $\chi_{j\mu} = \chi_j - \pi (\mu-1)/2$. On bipartite lattices, which we focus on in this letter, it is convenient to pick a gauge such that $e^{i\chi_j}$ is purely imaginary on one sublattice and purely real on the other one. The actual assignment of the phases is a matter of convenience and does not affect the physics. For simple cubic lattices a simple choice is $\chi_j = \bm{\pi} \cdot {\bf x}_j /2$, then
%
%With this choice 
the Majorana representation of $H$ in \eqref{eq:HubbardHamiltonian} becomes 
\begin{equation}
H = -\frac{t}{4} \sum_{\la i, j \ra,a}
e^{i(\chi_i - \chi_j)}\gamma_{i 1 a} \gamma_{j 1 a}
- \frac{U}{4} \sum_{i} 
\gamma_{i1\ma} \gamma_{i1\mb} \gamma_{i1\mc} \gamma_{i1\md}.
\label{eq:HubbardHamiltonianBetheMajorana}
\end{equation}
This form makes the well-known global SO(4) symmetry  \cite{Yang_Zhang_1990} of this model manifest; the six global SO(4) symmetry generators are
$Q_{ab} = \sum_{i,\mu} i \gamma_{i \mu a} \gamma_{i \mu b}$
($a > b$).

{\it Generators of canonical transformations.}
To restrict the class of canonical transformations we will in this letter only allow those that leave the theory SO(4)-symmetric and time-reversal invariant. We build these by considering a direct product of identical local transformations acting independently on every unit cell; in the following we will therefore momentarily drop the unit cell index when possible. The set of all the local canonical transformations is generated by the set of even Hermitean combinations of Majoranas $\{S_\alpha\}$ by exponentiation $V = e^{i \sum_\alpha \theta_\alpha S_{\alpha}/2}$ with real parameters $\theta_\alpha$ \cite{Bazzanella2014b}. Enforcing the symmetries demands that the generators satisfy $[S_\alpha, Q_{ab}]=0$ and $\Theta S_{\alpha}\Theta^{-1}=-S_{\alpha}$, with $\Theta$ the antiunitary time-reversal operator \cite{Sakurai_QM}. The standard choice of the action of the time-reversal operator for spin-1/2 fermions 
$\Theta c_{i\mu\ua} \Theta^{-1} = c_{i\mu\da}$ and 
$\Theta c_{i\mu\da} \Theta^{-1} = -c_{i\mu\ua}$
is realized on our Majoranas as
\begin{align}
\begin{split}
\Theta \gamma_{j\mu \ma} \Theta^{-1} &= s_{j\mu} \gamma_{j\mu \mc}
,\qquad
\Theta \gamma_{j\mu \mc} \Theta^{-1} =  - s_{j\mu} \gamma_{j\mu \ma},
\\
\Theta \gamma_{j\mu \md} \Theta^{-1} &= s_{j\mu} \gamma_{j\mu \mb}
,\qquad
\Theta \gamma_{j\mu \mb} \Theta^{-1} =  - s_{j\mu} \gamma_{j\mu \md},
\end{split}
\end{align}
with $s_{j\mu} = e^{i 2 \chi_{j\mu}} = \pm 1$.
In the following we will call operators that are SO(4)-symmetric  ``white''. For each pair of bands $\mu,\nu$ there is one white bilinear that we denote $H_{\mu \nu} = \sum_a i \gamma_{\mu a}\gamma_{\nu a}$. $H_{\mu \nu}$ is odd (even) under $\Theta$ when $\mu$ and $\nu$ are on the same (different) sublattice. From within each band only one quadrilinear can be formed; these are the ``band parity operators'' 
$P_\mu = \gamma_{\mu \ma} \gamma_{\mu \mb} \gamma_{\mu \mc} \gamma_{\mu \md}$ \cite{paritycomment}, which are white and even under $\Theta$. This implies that there are no allowed transformations if we restrict ourselves to a single band.
With two bands it is easy to spot two allowed generators
\begin{align}
S_1 = i H_{12} P_1  , \qquad
S_2 = i H_{12} P_2  .
\end{align}
They commute and one can check that there are no additional generators that are allowed with operators from only two bands, most easily using a computer algebra package \cite{Cliffordpackagearxiv}. Alternatively the number of generators are easily counted by enumerating the states that are allowed to mix \cite{supplemental}.
To illustrate the power of this approach we will work out the 2-band case in detail by performing a generic transformation with the generators $S_1$ and $S_2$, i.e., $V = e^{i(\theta_1 S_1 + \theta_2 S_2)/2}$. The transformed interaction term of \eqref{eq:HubbardHamiltonianBetheMajorana} follows directly from
\begin{multline}
V P_1 V^\dagger =  A_0 P_1 + A_1 H_{12}  + A_2 H_{12} P_1 P_2
\\
+A_3 \bigl( \bigl[H_{12}\bigr]^2 - 4\bigr) P_1 + A_4 P_2.
\end{multline}
$A_i$ are functions of the transformation parameters $\theta_1$ and $\theta_2$, explicit expressions are provided in the supplemental material \cite{supplemental}. As evident, the interaction term has become partly quadratic in the new fermions; this is a key feature of this approach.
To understand how the transformation acts on the hopping term in \eqref{eq:HubbardHamiltonianBetheMajorana} we work out (denoting $h_a = i \gamma_{1a} \gamma_{2a}$)
\begin{multline}
V \gamma_{1\ma} V^\dagger  =
\bigl[ B_1  +B_2 (h_\mb h_\mc+h_\mc h_\md+h_\md h_\mb)  
\\
+B_3 h_\ma P_2 + B_4 (h_\mb+h_\mc+h_\md) P_1 
\bigr]
\gamma_{1\ma} .
\label{eq:transformedgamma1}
\end{multline}
The $B_i$ are functions just like the $A_i$ \cite{supplemental}. Using this we generate the transformed Hamiltonian. The hopping term generally becomes a correlated hopping term. In mean-field this term can in principle generate non-local pairing terms (for example extended s-, p-, or d-wave) as well as spin-spin interaction terms, although we do not consider these possibilities further here.

{\it Trial Hamiltonian.} We now write down a quadratic trial Hamiltonian $\tilde{H}$ that we will use to approximate the transformed theory with. It can be represented as
\begin{align}
\tilde{H} = -\sum_{\la i,j \ra,\sigma}
a^{\dagger}_{i\mu\sigma}T_{\mu\nu} a^{}_{j\nu\sigma}
- \sum_{i,\sigma} a^{\dagger}_{i\mu\sigma}
\Lambda_{\mu\nu}
a^{}_{i\nu\sigma}.
\end{align}
The $a_{i\mu\sigma}$ are formed from the transformed Majoranas according to \eqref{eq:majoranadef}, $T_{\mu\nu}$ and $\Lambda_{\mu\nu}$ are matrices and the sums over $\mu,\nu$ are implied.
By construction this Hamiltonian is also manifestly SO(4)-symmetric when written out in terms of the transformed Majoranas if $T$ and $\Lambda$ connect different sublattices only. Broken symmetry states can obviously also be constructed. In the 2-band case there are three variational parameters in the trial Hamiltonian that we denote by $t_1, t_2, \lambda$, so that
\begin{align}
T = 
\begin{pmatrix}
t_1 & 0 \\ 0 & t_2
\end{pmatrix}
,\qquad
\Lambda =
\begin{pmatrix}
0 & \lambda \\ \lambda & 0
\end{pmatrix}.
\label{eq:2bandpmatrices}
\end{align}
Since $\tilde{H}$ is quadratic and translationally invariant it is easily diagonalized and all expectation values can be evaluated exactly using Wicks theorem; some details are provided in the supplemental material \cite{supplemental}. One particular result is the expression for the original local Green's function $G$ in terms of the local transformed ones $\tilde{G}$
\begin{align}
G_{11} = Z \tilde{G}_{11}
& + 4 B_3^2 \tilde{G}^3_{22}
+ 12 B_3 B_4 \tilde{G}_{22} (\tilde{G}^2_{21} + \tilde{G}^2_{12})
\nonumber \\
& + 12 B^2_4 \tilde{G}_{11} (\tilde{G}_{11} \tilde{G}_{22} 
+2 \tilde{G}_{12} \tilde{G}_{21})
\nonumber \\
& + 48 B_2^2 \bar{h}^2 \tilde{G}_{11} (\tilde{G}_{11} \tilde{G}_{22} - \tilde{G}_{12} \tilde{G}_{21})
\nonumber \\
& + 48 B_2^2 \tilde{G}_{11} (\tilde{G}_{11} \tilde{G}_{22} - \tilde{G}_{12} \tilde{G}_{21})^2 .
\label{eq:transformedlocalgreen0}
\end{align}
Here the indices denote the bands, and all components have the same imaginary time difference $\tau$, i.e., $\tilde{G}_{\mu\nu} = \tilde{G}_{\mu\nu}(\tau)$. The coefficient $Z = (B_1+3 B_2 \bar{h}^2)^2$ can be interpreted as a kind of quasiparticle weight.

{\it Variational study.}  By searching for local minima of the energy functional $\bar{E}(\theta_1,\theta_2,t_1,t_2,\lambda)$, which is the expectation value of $H$ per unit cell in the transformed trial ground state, we find two different insulating solutions that we denote by MI1 and MI2. This procedure can, if one so prefers, be implemented as a mean-field scheme for the parameters $t_1$, $t_2$, and $\lambda$. The energies of these solutions are plotted for $d=1$ in Fig.~\ref{fig:energyplot} together with the energies of a static antiferromagnetic mean-field solution, the exact result \cite{LiebWu1968}, and the result of the variational self-energy approach \cite{Potthoff2003} with the Hubbard-I self-energy \cite{HubbardI}. MI1 and MI2 cross around $U/t \approx 5.85$. Clearly both MI1 and MI2 get the leading term ($\sim - U/4$) correctly in the strong coupling limit, but neither MI1 nor MI2 are particularly good descriptions of the ground state since they do not have the correct spin correlations.
When the MI2 solution exists, the energy gain with respect to the local limit is to a very good approximation $-J d/4$, where $J = 4t^2/U$ is the exchange energy. This corresponds to the constant term generated in the mapping to the Heisenberg model \cite{Anderson1959}. The energy gain due to the spin correlations is therefore not captured by MI2. MI1, on the other hand,  does not capture any energy processes of order $J$ at strong coupling.
In $d=1$ MI2 only exist as a local minimum for $U \gtrsim 3.7$, while on the Bethe lattice MI1 does not exist for $U \gtrsim 4$.
\begin{figure}
\begin{center}
\includegraphics[scale=.6]{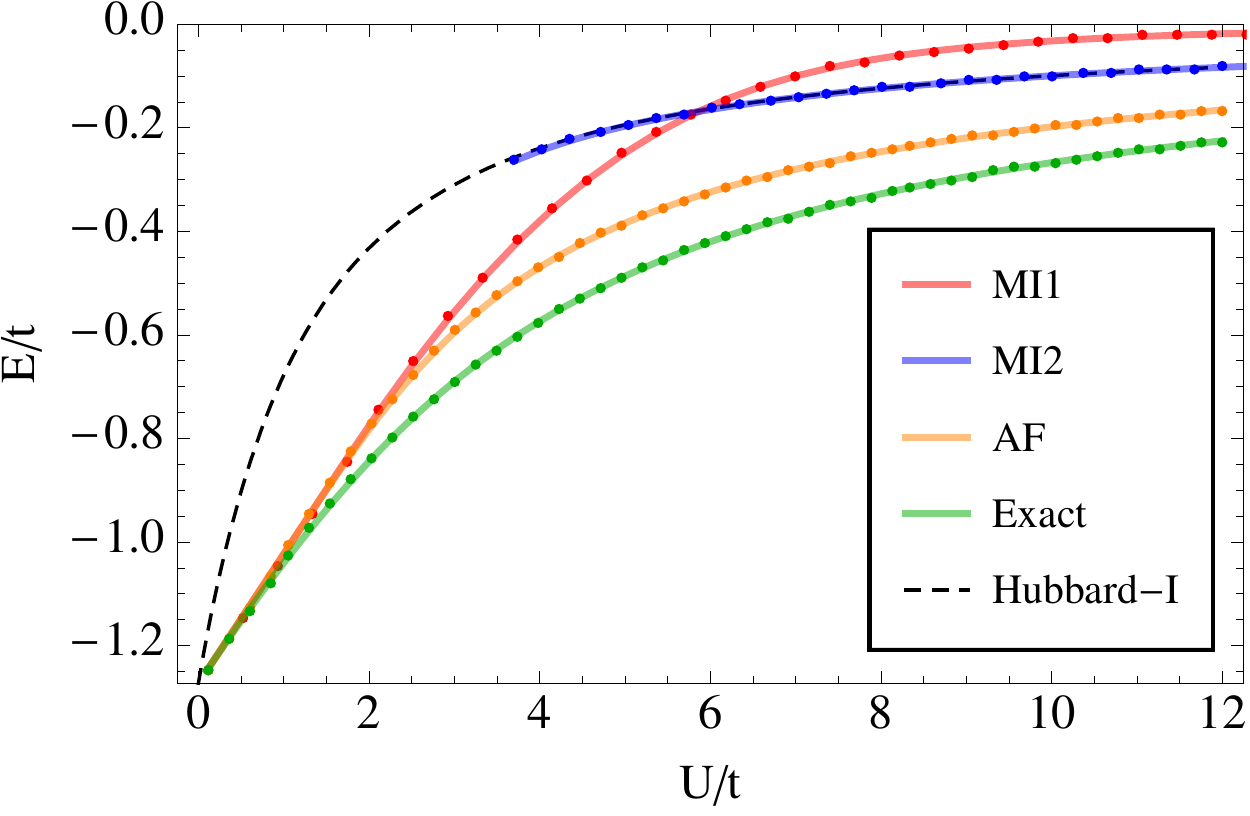}
\end{center}
\caption{Variational energies for different trial states in $d=1$: MI1, MI2, AF (static antiferromagnetic mean-field), exact, self-energy functional result with Hubbard-I self-energy.
The energy is shifted so that it goes to zero in the limit $U \rightarrow \infty$.
\label{fig:energyplot}}
\end{figure}

Both solutions are insulators, but for all practical purposes MI1 is metallic up to about $U \approx 1.7$ (for $d=1$) since the gap is exponentially small. Indeed, solving the mean field equations we find that the single particle gap is 
$\Delta \approx 16 t  \exp \bigl[-3 (1-4/\pi^2) (4 t/U)^2 \bigr]$ for small $U/t$ \cite{supplemental}. Similar results are obtained also for higher $d$, where a metallic state is expected for weak interactions. This tiny gap is the best that our 2-band system, which is generically gapped, can do to mimic a metal.

In Fig.~\ref{fig:zplot} we plot the quasiparticle weight $Z$ and the expectation value of the local parity $P = \la P_1 \ra$. $Z$ quantifies how much of the spectral weight of the original fermions that is described by free transformed fermions, c.f. \eqref{eq:transformedlocalgreen0}; $P$ instead is a measure of the amount of correlation in the states \cite{Gebhard_review}.
From the figure it is clear that $Z$ is always close to unity for MI2. For MI1 $Z$ starts out close to one for small $U/t$ and decreases as $U/t$ grows, as a result the excitations of the MI1 are largely incoherent for large and intermediate $U/t$. The correlation of both states is sizable for high and intermediate $U/t$, especially for MI2.
\begin{figure}
\begin{center}
\includegraphics[scale=.6]{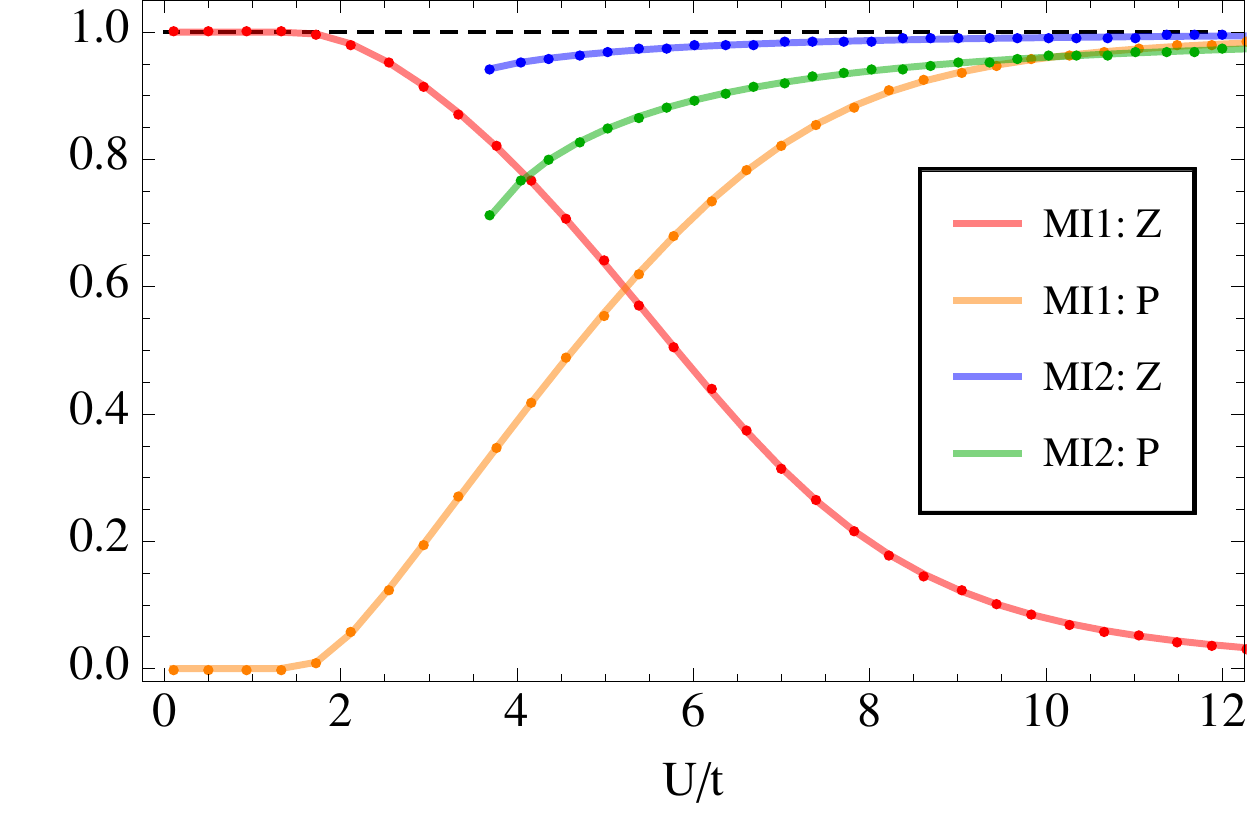}
\end{center}
\caption{Quasiparticle weight $Z = (B_1+3 B_2 \bar{h}^2)^2$ and band parity expectation value  
$P = -\la (2 n_{1\da}-1)(2 n_{1\ua}-1) \ra$ for the two states in $d=1$.
\label{fig:zplot}}
\end{figure}

The spin-spin correlation function of the original fermions can also be evaluated using Wicks theorem, and is always found to be small for MI2 \cite{supplemental}. This allows us to draw important conclusions about the nature of the MI2 state. When $P \approx 1$ there is one fermion per site, and since the spins are largely uncorrelated there is an extensive entropy of approximately $\ln 2$ per site in the physical subsystem. This pushes down the Free energy of the MI2 state below that of any fully ordered state for temperatures $T \gtrsim J $, as is expected for a PMI.

Some representative results for the spectral functions on the $d \rightarrow \infty$ Bethe lattice are plotted in Fig.~\ref{fig:spectralfunction}. For small $U/t$ MI1 starts out resembling the non-interacting case with a tiny gap. Increasing $U/t$ the gap grows and tails coming from the incoherent 3-particle contribution appear. MI2 mainly consists of two coherent Hubbard bands centered at energy $\approx \pm U/2$. In addition there are high-energy features around $\pm 3U/2$ and $\pm 5U/2$ with very small spectral weight (not shown). 
We would also like to stress that the  spectral weight sum rule for the fermion Green's function is obeyed exactly when all contributions to the spectral weight are taken into account \cite{supplemental}.

{\it Conclusions and future directions.} We have proposed a rather general method that can be used to generate correlated trial states (density matrices) that are potentially useful to understand and describe many strongly correlated systems. The method is easily generalized to treat broken symmetry states, systems away from half filling, and finite temperatures. In this work we have focused on the simplest description of a paramagnetic Mott insulator that it provides, which appears to be similar to the Hubbard-I approximation \cite{HubbardI} in some aspects, although our solution is obtained in a completely different way. Let us conclude with some comments about possible future applications of the method.

\begin{figure}
\begin{center}
\includegraphics[scale=.6]{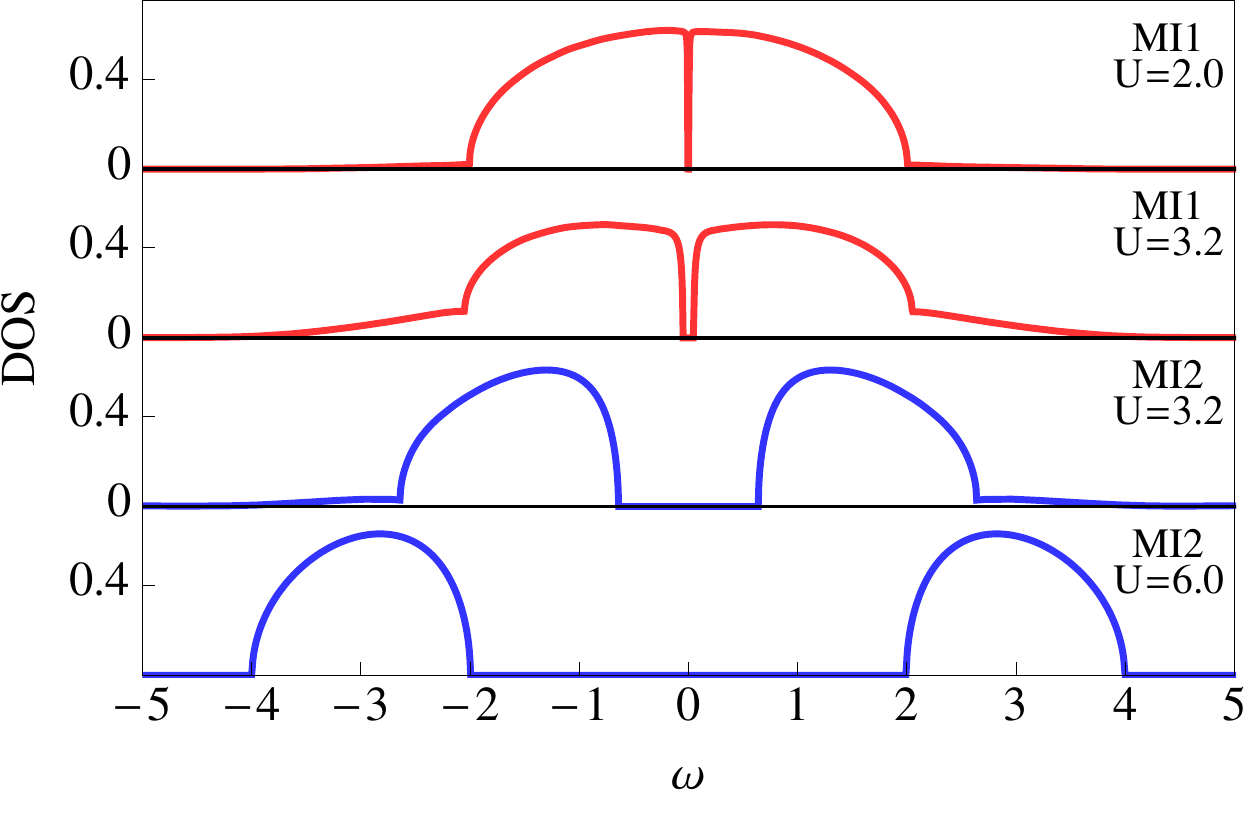}
\end{center}
\caption{Evolution of the local density of states on the $d \rightarrow \infty$ Bethe lattice with $U$. Energies are measured in units of $t$. On this lattice a first order transition between MI1 and MI2 takes place at $U \approx 3.2$ in the 2-band model.
\label{fig:spectralfunction}
}
\end{figure}

A description of the Mott metal-insulator transition at half filling within our scheme would, in the simplest setting, involve 3-band model, which generically leads to a metal. The allowed transformation becomes considerable more complex in this case since it in general involves 26 angles \cite{supplemental}. A preliminary study of a particular subclass of these transformations indicates that the insulating state becomes unstable to the presence of a third band for weak enough interactions \cite{nilssonunpublished}. This implies a paramagnetic metal-insulator transition that is consistent with the DMFT scenario \cite{DMFT_RMP1996}.

Although we do not consider doped Mott insulators in this work, we wish to remark that dynamical spectral weight transfer, which is a characteristic of Mott physics \cite{Phillips_RMP}, is straightforwardly captured within our multi-band scheme. This is already clear from Fig.~\ref{fig:spectralfunction} where there is a large rearrangement of the spectral weight at the transition between MI1 and MI2. Tuning the chemical potential $\mu$ in the gapless case the parameters of the non-linear canonical transformation will also change (away from half filling a generic local transformation in the 2-band model involves 5 angles). As a result the coupling of the physical fermions to the different bands changes. At half filling the coupling to the upper and lower Hubbard bands are equal. Going away from half filling by hole doping the coupling to the lower (upper) Hubbard band is expected to increase (decrease). This can be interpreted as dynamical spectral weight transfer.

Let us finally speculate about the consequences of the form of the transformed Green's function when the system is gapless. At low enough energies the spectral function will be dominated by the coherent single-particle contribution, even if $Z$ is small, as long as it is nonzero. This is consistent with the Landau Fermi liquid phenomenology. For intermediate energies it is certainly possible (in particular for small $Z$) that the spectral function is dominated by incoherent 3-particle excitations, which could possibly lead to strange metallic behavior. We leave a thorough investigation of this scenario for a future study.

\acknowledgments

We acknowledge useful discussions with Mats Granath and the Swedish research council (Vetenskapsr{\aa}det) for funding.
 
 \bibliography{MottMajoranaRefs}

\begin{thebibliography}{10}
\bibitem{paritycomment}%
  \BibitemOpen
  \bibinfo {note} {Actually there is minus sign with respect to the
  conventional parity since $P = -(2n_\ua -1) (2n_\da -1)$. This could be
  changed by a redefinition of the Majoranas.}%
  \bibAnnoteFile{Stop}{paritycomment}%
\end{thebibliography}

\section{Supplementary material}

In this supplemental material we provide some details and derivations that did not fit into the main text.

\section{Simple cubic lattices}

The effective lattice Hamiltonian is severely constrained if we assume that all symmetries [SO(4), time-reversal, and full lattice point group] are unbroken. On a generic lattice the trial Hamiltonian can then be written in Fourier space as
\begin{align}
\tilde{H} = -\sum_{\vk ,\sigma}
a^{\dagger}_{\vk\mu\sigma}T_{\mu\nu}(\vk) a^{}_{\vk\nu\sigma}
- \sum_{\vk,\sigma} 
a^{\dagger}_{\vk\mu\sigma}\Lambda_{\mu\nu} a^{}_{\vk\nu\sigma},
\end{align}
where the sums over the band indeces are implied, but suppressed.
In particular, for a fully symmetric simple cubic lattice in $d$ dimensions one has
\begin{equation}
%\begin{split}
T_{\mu\nu}(\vk) = 2 T_{\mu\nu} \chi_\vk,
\qquad
\chi_\vk =   \sum_{j=1}^d  \cos(\hat{\bf a}_j \cdot \vk).
%\end{split}
\end{equation}
In the 2-band model parametrized by the matrices in Eq.~\eqref{eq:2bandpmatrices}, there are two energy bands with energies
\begin{align}
E_{\pm}(\vk) = - (t_1+t_2)\chi_\vk 
\pm \sqrt{(t_1-t_2)^2\chi^2_\vk+\lambda^2} .
\end{align}
The system is gapless if there exists a $\vk$ such that $\chi_\vk = \chi_c \geq 0$ with $4 t_1 t_2 \chi_c^2 = \lambda^2$. Therefore the system is always gapped in $d$ dimensions when
\begin{align}
4 t_1 t_2 d^2 < \lambda^2 .
\end{align}
In the gapped case the $E_+$ band is empty and the $E_-$ band is filled in the ground state.  Averages can be expressed in a compact way using the density of states of the dimensionless energy $\chi$
\begin{align}
N(\chi) = \frac{1}{N}\sum_{\vk} \delta (\chi - \chi_\vk).
\end{align}
For simple cubic lattices in 1D and 2D the expressions are
\begin{align}
N_{1D}(\chi) &= \frac{1}{\pi \sqrt{1-\chi^2}} , \\
N_{2D}(\chi) &= \frac{1}{\pi^2}K(1- \chi^2 / 4) .
\end{align}
In many cases the results for the Bethe lattice in $d \rightarrow \infty$ with a proper (i.e., $1/\sqrt{d}$) rescaling of the hopping term can be obtained from the formulas for the simple cubic lattices by a substitution of the $N(\chi)$. In the Bethe lattice the support of $N_{Bethe}(\chi)$ is $[-1,1]$ and 
\begin{align}
N_{Bethe}(\chi) &= \frac{2}{\pi}\sqrt{1-\chi^2} .
\label{eq:dosBethe}
\end{align}

\subsection{Trial Green's functions}

In Fourier space the trial Matsubara Green's function (dropping the diagonal spin index) is given by
\begin{align}
\tilde{G}_{\vk}(\io)
=
\begin{pmatrix}
\io+ 2 t_1 \chi_\vk & \lambda \\
\lambda & \io+ 2 t_2 \chi_\vk
\end{pmatrix}^{-1}.
\end{align}
Inverting this matrix we get the explicit expression
\begin{multline}
\tilde{G}_{\vk}(\io) = 
\frac{1}{2}
\Bigl[ \frac{1}{\io-E_{-}(\vk)} + \frac{1}{\io-E_{+}(\vk)} \Bigr] \mathbb{1}
\\
+ 
\frac{1}{2 F_\vk }
\Bigl[ \frac{1}{\io-E_{-}(\vk)} - \frac{1}{\io-E_{+}(\vk)} \Bigr]
\begin{pmatrix}
t_{12} \chi_\vk & \lambda \\
\lambda & - t_{12} \chi_\vk
\end{pmatrix},
\label{eq:greensfunctionk}
\end{multline}
where
\begin{align}
F_\vk = \sqrt{ t_{12}^2\chi^2_\vk+\lambda^2},
\qquad
t_{12 } = t_1 -t_2.
\end{align}
Given the Green's function all averages can be calculated using standard machinery, see e.g. \cite{mahan}.

\subsection{Majorana Green's functions}

The normal fermion Green's functions and those of the Majoranas are related in a simple way. In particular the imaginary time real space Green's function is \cite{mahan}
\begin{align}
\tilde{G}_{i\mu;j\nu}(\tau) &=  - \la T_\tau  a^{}_{i \mu \sigma}(\tau)  a^\dagger_{j \nu \sigma}(0)\ra
\nonumber \\
&= -\frac{e^{i (\chi_{j\nu} -\chi_{i\mu})}}{2}
 \la T_\tau \gamma_{i\mu a}(\tau) \gamma_{j\nu a}(0)  \ra ,
\end{align}
where the phases $\chi_{i\mu}$ where introduced after Eq.~\eqref{eq:majoranadef}.
For the (transformed) Majoranas we define the corresponding Green's function via
\begin{align}
\tilde{g}_{i\mu;j\nu}(\tau) = -
 \la T_\tau \gamma_{i\mu a}(\tau) \gamma_{j\nu a}(0)  \ra .
 \label{eq:majoranagreendef}
\end{align}
Note that since $\tilde{G}_{i1;j2} = \tilde{G}_{i2;j1}$ we have 
$\tilde{g}_{i1;j2} = -\tilde{g}_{i2;j1}$. For the local Green's function we have $\tilde{g}_{i\mu;i\mu} = 2\tilde{G}_{i\mu;i\mu}$ and and $\tilde{g}_{i1;i2} = -\tilde{g}_{i2;i1} = 2i \tilde{G}_{i1;i2}$.

\subsection{Determination of the parameters -- \\Variational mean field theory}

Following the scheme developed in unpublished work by \"Ostlund (see also Appendix of \cite{Bazzanella2014a}) we introduce a trial Hamiltonian
$\tilde{H} = \sum_i \tilde{\mu}_i A_i$
and the corresponding trial density matrix $\tilde{\rho} \propto e^{-\beta \tilde{H}}$. Denoting $\alpha_i = \text{Tr} A_i \tilde{\rho}$ the variational parameters that extremize (minimize) the trial Free energy are given by
\begin{equation}
\tilde{\mu}_i = \frac{\partial}{\partial \alpha_i} \text{Tr} H \tilde{\rho} .
\end{equation}
An important consequence of this equation is that only terms that are generated in the Wick expansion can be present in the mean field theory.

\section{Canonical transformation -- \\ 2-band calculational details}

Let us consider the transformation generated by $V= e^{i (\theta_1 S_1 +\theta_2 S_2)/2}$ and denote $h = H_{12}$. It is now straightforward to calculate the transformed Hamiltonian, using for example
\begin{align}
V P_1 V^\dagger = V^2 P_1 .
\end{align}
Working out the algebra we find that (recall the definition $h_a = i \gamma_{1a} \gamma_{2a}$)
\begin{multline}
V^2  = A_0 + A_1 h P_1 + A_2 h P_2
\\
+A_3 (h_\ma h_\mb + \text{permutations}) + A_4 P_1 P_2.
\end{multline}
The functions are given by the following expressions
\begin{subequations}
\begin{align}
A_0 &=  
\frac{\cos(4 \theta_1 + 4 \theta_2)+4 \cos(2\theta_1-2\theta_2)+3}{8}   ,
\\
A_1 &=  
\frac{-\sin(4 \theta_1 + 4 \theta_2)-2\sin(2\theta_1-2\theta_2)}{8},
%- \frac{\sin(2\theta_1-2\theta_2)}{4} ,
\\
A_2 &= 
\frac{-\sin(4 \theta_1 + 4 \theta_2)+2\sin(2\theta_1-2\theta_2)}{8},
%-\frac{\sin(4 \theta_1 + 4 \theta_2)}{8} + \frac{\sin(2\theta_1-2\theta_2)}{4} ,
\\
A_3 %= -\sin^2(\theta_1+\theta_2) \cos^2(\theta_1+\theta_2)
&= %-\frac{\sin^2(2 \theta_1 + 2 \theta_2)}{4}
\frac{\cos(4 \theta_1 + 4 \theta_2)-1}{8} ,
\\
A_4 &= %& \sin^4(\theta_1+\theta_2)+\sin^4(\theta_1-\theta_2) \nonumber \\
% -& \sin(2\theta_1) \sin(2\theta_2) [\cos^2(\theta_1+\theta_2)+\cos^2(\theta_1-\theta_2)]
%\nonumber \\
 % \frac{\cos(4 \theta_1 + 4 \theta_2)}{8} - \frac{\cos(2\theta_1-2\theta_2)}{2} + \frac{3}{8} 
  \frac{\cos(4 \theta_1 + 4 \theta_2)-4 \cos(2\theta_1-2\theta_2)+3}{8}  .
\end{align}
\end{subequations}
The calculation of the transformed hopping term is also straightforward and the result is presented in the main text in Eq.~\eqref{eq:transformedgamma1}, rewritten here for convenience
\begin{align}
V \gamma_{1\ma} V^\dagger  &=
\bigl[ B_1  +B_2 (h_\mb h_\mc+h_\mc h_\md+h_\md h_\mb)  
\nonumber \\
&+B_3 h_\ma P_2 + B_4 (h_\mb+h_\mc+h_\md) P_1 
\bigr]
\gamma_{1\ma} .
\label{eq:transformedgamma1app}
\end{align}
where
\begin{subequations}
\begin{align}
B_1  &= \frac{\cos(3 \theta_1 + \theta_2) + 3 \cos(\theta_1 - \theta_2)}{4},
\\
B_2 &= \frac{\cos(3 \theta_1 + \theta_2)-\cos(\theta_1 - \theta_2)}{4} ,
\\
B_3 &= \frac{-\sin(3 \theta_1 + \theta_2)+3 \sin(\theta_1 - \theta_2)}{4},
\\
B_4
&= 
\frac{-\sin(3 \theta_1 + \theta_2)-\sin(\theta_1 - \theta_2)}{4} .
\end{align}
\end{subequations}

\section{Averages in the 2-band model}

To perform the variational calculation we need to compute the averages that appear in the Wick decomposition of the Hamiltonian. In the 2-band model there is one local average and two non-local ones. In this section we denote the Majoranas on two neighboring units cells with and without a prime. The averages are
\begin{subequations}
\begin{align}
\bar{h} &= \la   i \gamma_{1a} \gamma_{2a} \ra
=\la a_{1\sigma}^\dagger a^{}_{2\sigma} + \text{h.c.} \ra ,
\\
\bar{k}_1 &= e^{i(\chi - \chi')} \la \gamma_{1a} \gamma'_{1a} \ra  
= \la a_{1\sigma}^\dagger a'_{1\sigma} + \text{h.c.} \ra ,
\\
\bar{k}_2 & = e^{i(\chi - \chi')} \la \gamma_{2a} \gamma'_{2a} \ra 
= \la a_{2\sigma}^\dagger a'_{2\sigma} + \text{h.c.} \ra .
\end{align}
\end{subequations}
In writing these expressions we have used the flavor and spin symmetries. The bond averages are also independent of their direction since we have assumed full lattice symmetry. To avoid confusion we stress that these expressions hold for the averages, not on the operator level.
The average of the local term in the Hamiltonian is
\begin{align}
-\frac{U}{4}\la V P_1 V^\dagger \ra  &=
-U (A_1 \bar{h} + A_2 \bar{h}^3) ,
\end{align}
while the average of the hopping term becomes
\begin{multline}
 \bar{b} = e^{i(\chi - \chi')} \la V  \gamma^{\,}_1 \gamma_1' V^\dagger \ra  =
(B_1+3 B_2 \bar{h}^2)^2 \bar{k}_1
\\
+ 3 B_2^2 \bar{k}_1^2 \bar{k}_2 (\bar{k}_1 \bar{k}_2 - 4 \bar{h}^2)
- \bar{k}_2 (B_3^2 \bar{k}_2^2 + 3 B_4^2 \bar{k}_1^2) .
\end{multline}
The energy functional to minimize at $T=0$, which is nothing but energy expectation value per unit cell, is
\begin{align}
\bar{E} = \frac{\la V H V^\dagger \ra}{N}
=
-t 2d \bar{b} -  U (A_1 \bar{h} + A_2 \bar{h}^3).
\end{align}
The factor 2 in the first term is due to spin. The variational mean field theory introduced above implies that the parameters of the trial Hamiltonian are given by
\begin{align}
\lambda = -\frac{1}{2} \frac{\partial \bar{E}}{\partial \bar{h}}
, \qquad
t_\mu = -\frac{1}{2d} \frac{\partial \bar{E}}{\partial \bar{k}_\mu} .
\end{align}
Explicitly the resulting expressions are
\begin{subequations}
\begin{multline}
t_1  = t \Bigl[
(B_1+3 B_2 \bar{h}^2)^2 + 9 B_2^2 \bar{k}^2_1 \bar{k}^2_2
\\
- 6 \bar{k}_1 \bar{k}_2 (B_4^2 + 4 B_2^2 \bar{h}^2) \Bigr],
\end{multline}
\begin{align}
t_2  = -3 t  \Bigl[ B_3^2  \bar{k}^2_2 + B_4^2  \bar{k}^2_1
+4 B_2^2 \bar{k}^2_1 \bar{h}^2
-2 B_2^2 \bar{k}^3_1 \bar{k}_2 \Bigr],
\end{align}
\begin{multline}
\lambda = \frac{U}{2} (A_1 + 3 A_2 \bar{h}^2)
\\
+ 12 d t  \bar{k}_1 B_2 \bar{h} (B_1 + 3 B_2 \bar{h}^2 -
2 B_2 \bar{k}_1 \bar{k}_2).
\end{multline}
\end{subequations}
Clearly, to minimize the expectation value of the kinetic term (recall that $t > 0$) it is best if $\bar{k}_1 \geq 0$ and $\bar{k}_2 \leq 0$. This in turn implies that in the mean field equations $t_1 \geq 0$ and $t_2 \leq 0$, and therefore the system is always gapped if $\lambda \neq 0$. At $T=0$ there is also the relation $\bar{k}_2 = -\bar{k}_1$ which follows from the form of the Green's function in Eq.~\eqref{eq:greensfunctionk}.
Let us specialize to this case, denoting $\bar{k} = \bar{k}_1$ and setting $t_{12} = t_1 - t_2$. The averages can then easily be obtained from Eq.~\eqref{eq:greensfunctionk} with the result
\begin{subequations}
\label{eq:averageintegrals1}
\begin{align}
\bar{h} &= \int^d_{-d} d\chi N(\chi) \frac{\lambda}{\sqrt{t_{12}^2 \chi^2 + \lambda^2}},
%\end{align}
\\
%\begin{align}
\bar{k} &= \frac{t_{12}}{d} 
\int^d_{-d} d\chi N(\chi) \frac{\chi^2}{\sqrt{t_{12}^2 \chi^2 + \lambda^2}}.
\end{align}
\end{subequations}
Note that the magnitudes of these only depend on the quantity $r = t_{12}/\lambda$, thus $\bar{E} = \bar{E}(\theta_1,\theta_2,r)$. This implies that minimization of the energy functional typically determines the values of three of the variational parameters $\theta_1$,  $\theta_2$, and $r$. The value of $r$ can also be obtained by solving the mean field equations for $t_1$, $t_2$ and $\lambda$. Mean field theory therefore provides one consistent way of determining the parameters of the trial Hamiltonian; but it is not the only possible scheme that can be considered.

The transformation angles for $d=1$ are presented in Fig.~\ref{fig:angleplot}. MI1, which is connected with the non-interacting limit, is best for small $U/t$. It starts out with small values of the transformation angles for small $U/t$, while for larger values of $U/t$ the angles saturate at $\theta_1 = -3 \pi/16$ and $\theta_2 = \pi/16$. MI2, which is good for large $U/t$ has values of the angles that are close to $\theta_1 = \pi/16$ and $\theta_2 = -3 \pi/16$, which is appropriate for the local limit.
\begin{figure}
\begin{center}
\includegraphics[scale=.6]{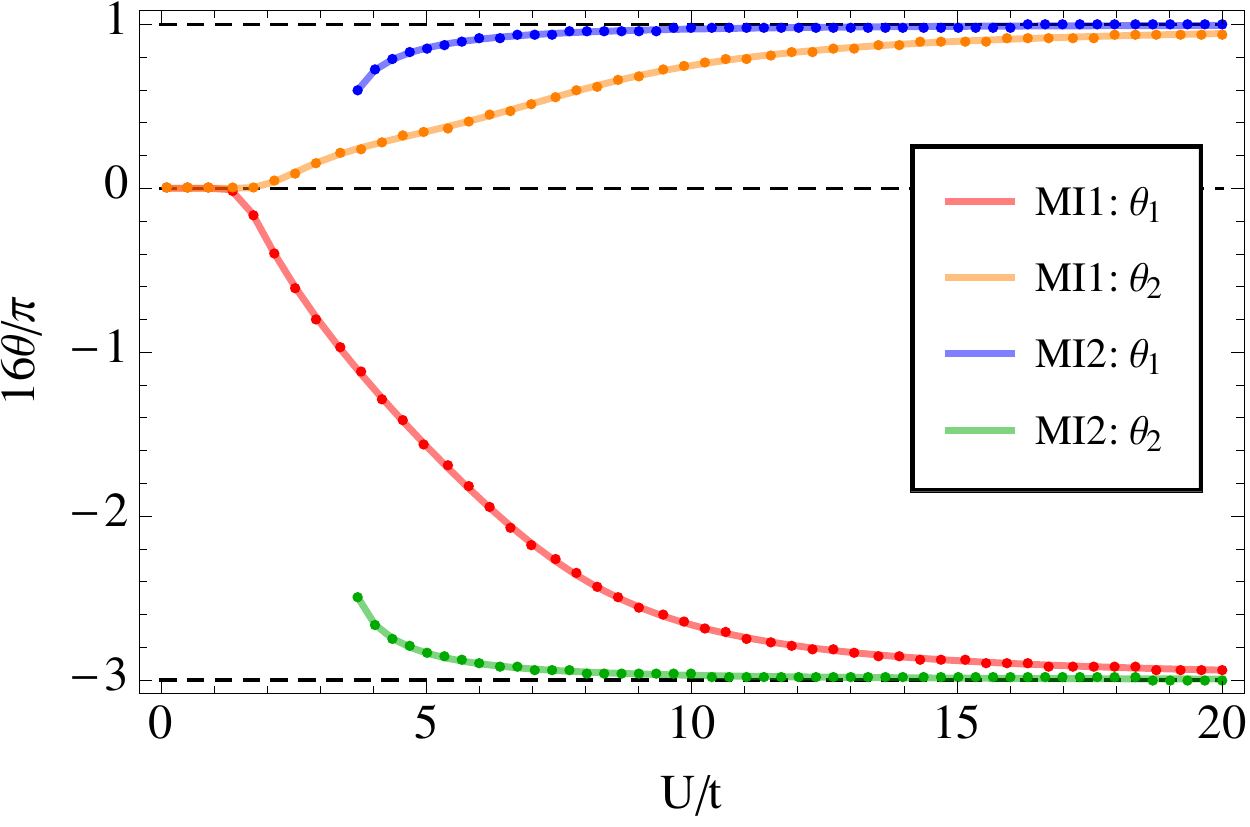}
\end{center}
\caption{Evolution of the transformation angles $\theta_1$ and $\theta_2$ for the two Mott insulating states in $d=1$.
\label{fig:angleplot}}
\end{figure}

In the following subsections we analyze a few special cases where we can get partly analytical results.

\subsection{Local limit}

To get the energy of the local problem correctly one may set use a trial Hamiltonian $\tilde{H} = -\lambda H_{12}/2$, and assuming $\lambda>0$ one gets $\bar{h}=1$ and to minimize the energy one should set $\sin(4\theta_1+4\theta_2)=-1$. The trial energy is then independent of the angle difference $\theta_2-\theta_1$. To fix $\lambda$ we may use mean field theory which leads to
\begin{align}
\lambda &= \frac{U}{2} (A_1+3 \bar{h}^2 A_2)
\nonumber \\
& = \frac{U}{2} \frac{-\sin(4 \theta_1 + 4 \theta_2) + \sin(2\theta_1-2\theta_2)}{2}
\nonumber \\
&= \frac{U}{2} \frac{1 + \sin(2\theta_1-2\theta_2)}{2}.
\label{eq:localmeanfieldlambda}
\end{align}
We may maximize this by picking $ \sin(2\theta_1-2\theta_2) = 1$. A possible choice is $\theta_1 = \pi/16$ and $\theta_2 = -3\pi / 16$, which corresponds to the asymptotic behavior of MI2.
Let us now consider a state where $h_a |\Omega \ra = \bar{h} |\Omega \ra$. When acting on such a state we immediately have that
\begin{multline}
V \gamma_{1\ma} V^\dagger  |\Omega \ra 
\\ =
(B_1 + 3 B_2 \bar{h}^2) \gamma_{1\ma}  |\Omega \ra
-P_1 (B_3 \bar{h}^3 + 3 B_4 \bar{h})\gamma_{1\ma}  |\Omega \ra.
\label{eq:transformedgammalocal}
\end{multline}
With our choice of angles and using the fact that $\bar{h}=1$ this gives
\begin{equation}
V \gamma_{1\ma} V^\dagger  |\Omega \ra = \gamma_{1\ma}  |\Omega \ra .
\label{eq:localgammaaction}
\end{equation}
This implies that the transformed $\gamma_{1\ma}$ only creates single-particle excitations on top of the new vacuum. The second term in Eq.~\eqref{eq:transformedgammalocal} will for general angles create a three-particle excitations. In particular, when the angles are exchanged so that $\theta_1 = -3\pi/16$ and $\theta_2 = \pi / 16$, which corresponds to the asymptotic behavior of MI1, one finds that
$V \gamma_{1\ma} V^\dagger  |\Omega \ra =\gamma_{1\mb} \gamma_{1\mc} \gamma_{1\md}  |\Omega \ra$.

Using Eq.~\eqref{eq:localgammaaction} it is straightforward to calculate the Green's function for the Majoranas. Using $\mathcal{O}(\tau) = e^{\tilde{H} \tau} \mathcal{O} e^{- \tilde{H} \tau}$ with $\lambda = U/2$ (corresponding to MI2 in the strong coupling limit) we get
\begin{equation}
g_{11}(\io) = \tilde{g}_{11}(\io)
= \frac{1}{\io - U/2} + \frac{1}{\io + U/2},
\end{equation}
which exactly reproduce the correct Green's function in the local limit. The exact local Green's function can also be obtained from the angles corresponding to MI1 if one sets $\lambda = U/6$; this is however not possible using a mean field decoupling scheme since according to Eq.~\eqref{eq:localmeanfieldlambda} $\lambda = 0$ for the angles corresponding to MI1.

\subsection{Large $U/t$ limit}

For large $U/t$ we expect $r = t_{12}/\lambda$ to be small. In this case we have (assuming $\lambda > 0$)
\begin{align}
\bar{h} \approx 1 -  r^2 I_2/2 , \qquad \bar{k} \approx r I_2 / d .
\end{align}
Here we introduced the integral $I_2 = \int^d_{-d} d\chi N(\chi) \chi^2 $, which for simple cubic lattices in $d$ dimensions evaluates to $I_2 = d/2$. To leading order in $r$ the variational energy becomes
\begin{multline}
\bar{E} \approx
-U (A_1  + A_2)
+ U (A_1 +3 A_2 ) I_2  r^2 /2
\\
 - 2 t (B_1+3 B_2)^2 I_2 r .
\end{multline}
Viewed as a function of $\alpha = 4(\theta_1+\theta_2)$ and $\beta = 2(\theta_1-\theta_2)$ this function always have extrema at $\alpha,\beta = \pm \pi/2$; two of these combinations have low energies. One with $\alpha = \beta = -\pi/2$ has energy $-U/4$. The other one, corresponding to MI2, has $\alpha = -\beta = -\pi/2$ and its energy is
\begin{equation}
\bar{E} = -\frac{U}{4} + \frac{U I_2 r^2}{2} - 2 I_2 r t.
\end{equation}
Minimizing this we find $r = 2t/U$ and the energy $\bar{E} = -U/4- 2 I_2 t^2/U$.

In the large $U$ limit the Hamiltonian can be mapped onto a Heisenberg model \cite{Anderson1959}
\begin{equation}
H  \rightarrow -N \frac{U}{4} + H_J , \qquad
H_J = J \sum_{\la i,j\ra}({\bf S}_i \cdot {\bf S}_j - \frac{1}{4}),
\end{equation}
with $J = 4 t^2/U$. We can therefore conclude that in the strong coupling limit the energy of the MI2 state is that of the Heisenberg model without any nearest neighbor spin-spin correlations.

\subsection{Small $U/t$ limit}

In this limit we expect $\delta = r^{-1} = \lambda/t_{12}$ to be small. We then have (assuming $\lambda , t_{12} > 0$) that
\begin{subequations}
\begin{align}
\bar{h} = \delta \int^d_{-d} d\chi N(\chi) \frac{1}{\sqrt{ \chi^2 +\delta^{2}}},
\end{align}
\begin{align}
\bar{k} = \frac{1}{d} 
\int^d_{-d} d\chi N(\chi) \frac{\chi^2}{\sqrt{\chi^2 + \delta^{2}}}.
\end{align}
\end{subequations}
In 1D this implies that
\begin{subequations}
\begin{align}
\bar{h} &\approx \frac{2\delta \ln(4/ \delta)}{\pi} ,
\\
\bar{k} &\approx \frac{2}{\pi}
+ \frac{ \delta^{2 } [1-2\ln(4/\delta)] }{2 \pi},
\end{align}
\end{subequations}
while on the Bethe lattice in $d \rightarrow \infty$
\begin{subequations}
\begin{align}
\bar{h} & \approx \frac{4 \delta}{\pi} \bigl[ \ln(4/\delta)-1 \bigr] ,
\\
\bar{k} & \approx \frac{1}{\sqrt{d}} 
\Bigl( \frac{4}{3\pi} +\frac{\delta^2}{\pi}[3-2 \ln(4/\delta)] \Bigr) .
\end{align}
\end{subequations}
The expansion on the simple cubic lattice in $d=2$ is more cumbersome because of the van Hove singularity.

In addition we know that the transformation angles are also small in this limit from the numerical solution of MI1. Using this we may expand the energy functional in $\delta$, $\theta_1$ and $\theta_2$; the resulting expression becomes
\begin{align}
\bar{E} \approx -2 t d \bar{k} 
+ 2 t d \bar{k} (1-\bar{k}^2) (3 \theta_1^2 + \theta_2^2)
+  U \bar{h} \theta_1 .
\end{align}
A similar expansion of the mean field equations gives
\begin{subequations}
\begin{align}
t_{12} &\approx t\bigl[ 1 -(3 \theta_1^2 + \theta_2^2) (1-3 \bar{k}^2) \bigr],
\\
\lambda &\approx - U \theta_1 /2 ,
\end{align}
\end{subequations}
so that $\delta \approx - U \theta_1 / (2t)$. Since $\bar{k}^2 < 1$ we may set $\theta_2 = 0$ and it remains only to minimize $\bar{E}$ with respect to $\theta_1$ (or $\delta$).
On the Bethe lattice
\begin{align}
\bar{E} \approx -\frac{8 t}{3 \pi} + \frac{2\delta^2}{\pi}
\bigl[ 1 + (4t/U)^2 -2\ln(4/\delta) \bigr]t ,
\end{align}
so that
\begin{align}
\theta_1 \approx - 8(t/U) \exp\bigl[ -1-8 (t/U)^2 \bigr],
\end{align}
therefore the resulting single particle gap is (since $t_2 = 0$)
\begin{multline}
\Delta = 2 \Bigl( \sqrt{t_1^2+\lambda^2}- t_1 \Bigr)
%\approx \frac{\lambda^2}{t_1}
%\\
\approx 16 t  \exp \bigl[-2 - (4 t/U)^2 \bigr].
\end{multline}

On the simple cubic lattice in $d=1$ a similar calculation gives
\begin{align}
\bar{E} \approx -\frac{4t}{\pi} + \frac{\delta^2}{\pi}
\bigl[ 48\Bigl(1-\frac{4}{\pi^2}\Bigr)(t/U)^2 - 1 -2\ln(4/\delta) \bigr]t ,
\end{align}
so that
\begin{align}
\theta_1 \approx - 8(t/U) \exp\bigl[ -24 (1-4/\pi^2)(t/U)^2 \bigr],
\end{align}
and hence the single particle gap is (since $t_2 \approx 0$)
\begin{multline}
\Delta = 2 \Bigl( \sqrt{t_1^2+\lambda^2}- t_1 \Bigr)
%\approx \frac{\lambda^2}{t_1}
%\\
\approx 16 t  
\exp\bigl[ -48 (1-4/\pi^2)(t/U)^2 \bigr] .
%\exp \bigl[-3 (4 t/U)^2 \bigr].
\end{multline}
We also obtain $\bar{E} \approx -4t/\pi - \Delta/\pi$. The asymptotic expressions in this subsection are found to be a good approximation to the numerical solutions also for moderate values of $U/t$.

\section{Green's functions}

It is just a matter of some straightforward but tedious algebra to work out the Green's function of the original operators in terms of the Green's functions of the transformed ones. For brevity we will suppress the site and time indeces, i.e., we use the shorthand notation $\tilde{g}_{\mu\nu} = \tilde{g}_{i\mu;j\nu} (\tau)$ where the first index goes with $i,\tau$ and the second with $j,\tau=0$ [see definition in Eq.~\eqref{eq:majoranagreendef}]. Using Eq.~\eqref{eq:transformedgamma1app} and Wicks theorem the original Majorana Green's function becomes the following polynomial in terms of the new ones
\begin{align}
g_{i1;j1}(\tau) = g_{11} &= Z \tilde{g}_{11}
\nonumber \\
& + 12 B_2^2 \bar{h}^2 \tilde{g}_{11} (\tilde{g}_{11} \tilde{g}_{22} - \tilde{g}_{12} \tilde{g}_{21})
\nonumber \\
& + 3 B_2^2 \tilde{g}_{11} (\tilde{g}_{11} \tilde{g}_{22} - \tilde{g}_{12} \tilde{g}_{21})^2
\nonumber \\
& + B_3^2 \tilde{g}^3_{22}
\nonumber \\
& - 3 B_3 B_4 \tilde{g}_{22} (\tilde{g}^2_{21} + \tilde{g}^2_{12})
\nonumber \\
& + 3 B^2_4 \tilde{g}_{11} (\tilde{g}_{11} \tilde{g}_{22} +2 \tilde{g}_{12} \tilde{g}_{21}).
\label{eq:transformedgreen}
\end{align}
Recall that $Z = (B_1+3 B_2 \bar{h}^2)^2$, as stated in the main text. This implies that the total Green's function consists of a coherent part and an incoherent part built up from 3- and 5-quasiparticle excitations.

\subsection{Incoherent contributions}

The frequency dependence of the incoherent contributions is easily obtained working with the imaginary time Green's functions in the Lehmann represenation, as we outline in this subsection. A generic component of a Matsubara Green's function $g(\io)$ can be written as
\begin{align}
g(\io) = \int d\e \frac{A(\e)}{\io - \e},
\label{eq:lehmann1}
\end{align}
where $A(\e)$ is a real weight function, which is not necessarily positive. In the imaginary time interval $-\beta < \tau < \beta$ the corresponding imaginary time Green's function becomes [$n_\e = 1/(1+e^{\beta \e})$] \cite{mahan}
\begin{align}
g (\tau) = \int d\e A(\e) e^{-\e \tau} \bigl[ n_\e-\Theta(\tau) \bigr].
\end{align}
This implies that the Matsubara Green's function corresponding to a product of three different imaginary time Green's functions $g_1 (\tau)g_2 (\tau)g_3 (\tau)$ becomes
\begin{multline}
[g_1 g_2 g_3 ](\io)  = 
\int d\e_1 d\e_2 d\e_3 A_1 (\e_1) A_2 (\e_2) A_3 (\e_3)
\\ \times
\frac{n_{\e_1}n_{\e_2}n_{\e_3}
+(1-n_{\e_1})(1-n_{\e_2})(1-n_{\e_3})
}{\io - (\e_1 + \e_2 + \e_3)}.
\end{multline}
This formula is easily generalized to products of higher order. To get the corresponding spectral function we make the usual replacement $1/(\io -\e) \rightarrow \delta(\omega - \e)$, so that
\begin{multline}
-\frac{1}{\pi} \text{Im} [g_1 g_2 g_3 ](\omega + i 0^+ ) 
  = 
\int d\e_2 d\e_3 A_1 (\e_1) A_2 (\e_2) A_3 (\e_3)
\\ \times
\bigl[ n_{\e_1}n_{\e_2}n_{\e_3}
+(1-n_{\e_1})(1-n_{\e_2})(1-n_{\e_3}) \bigr]_{\e_1 = \omega - \e_2 - \e_3}.
\end{multline}

\subsection{Sum rule check}
It can be demonstrated that the local Green's function in Eq.~\eqref{eq:transformedgreen} satisfies the spectral weight sum rule $-\frac{1}{\pi}\int d\omega \text{Im} g_{i1;i1}(\omega+i0^+) = 2$. Using the symmetries $\tilde{A}_{11}(-\e) = \tilde{A}_{11}(\e)$, $\tilde{A}_{22}(-\e) = \tilde{A}_{22}(\e)$, $\tilde{A}_{12}(-\e) = -\tilde{A}_{12}(\e)$ and the formulas above we find that (for any temperature) it is allowed to substitute $\tilde{g}_{11} = \tilde{g}_{22} = 1$ and $\tilde{g}^2_{12} = - \bar{h}^2$ everywhere and include an overall factor of 2. The resulting expression is always equal to 2 because of the structure of the $B$'s. This is a consequence of that the transformation is canonical and provides a useful consistency check on the theory.

\section{Spin correlations}

The spin operators are also easy to work out. We will drop the site index for brevity in this section, consider for example
\begin{align}
S_{1z} = \frac{i\gamma_{1\mb}\gamma_{1\ma}-i\gamma_{1\md}\gamma_{1\mc}}{4} .
\end{align}
Let us define symmetric and antisymmetric combinations
\begin{align}
B^{\pm}_{ab} = i\gamma_{1a}\gamma_{1b} 
\pm i \gamma_{2a}\gamma_{2b},
\end{align}
so that
\begin{align}
S_{1z} 
 =\frac{S_{1z}+ S_{2z}}{2} 
 + \frac{B^{-}_{\mb \ma} - B^{-}_{\md \mc}}{8} .
\end{align}
Transforming this we have
\begin{align}
V S_{1z} V^\dagger 
=
\frac{S_{1z}+ S_{2z}}{2} 
 + \frac{V B^-_{\mb \ma}V^\dagger - V B^-_{\md \mc} V^\dagger}{8} ,
\end{align}
since the transformation conserves total spin. Working out the transformation we find
\begin{multline}
V B^-_{\mb \ma} V^\dagger
= \bigl[ C_1 + C_2 h_\mc h_\md \bigr] B^-_{\mb \ma}
\\
+
(h_\ma+h_\mb) ( C_3 B^-_{\md \mc} + C_4 B^+_{\md \mc}),
\end{multline}
where
\begin{subequations}
\begin{align}
C_1  &= \frac{\cos(2 \theta_1 + 2\theta_2) + \cos(2\theta_1 -2 \theta_2)}{2},
\\
C_2 &= \frac{\cos(2 \theta_1 + 2\theta_2) - \cos(2\theta_1 -2 \theta_2)}{2} ,
\\
C_3 &= \frac{\sin(2 \theta_1 + 2\theta_2)}{2},
\\
C_4 &= \frac{\sin(2 \theta_1 - 2\theta_2)}{2} .
\end{align}
\end{subequations}
The formula for the transformed $B^-_{\md \mc}$ is analogous with the exchange $\ma,\mb \leftrightarrow \mc,\md$.
This implies that averages of spin correlation functions can be factorized using Wicks theorem with the result
\begin{multline}
\la V S_{1z} S'_{1z} V^{\dagger} \ra 
= \frac{1}{32}
\\
\times \sum_{s,s'=\pm}
\la B_{\mb \ma}^s {B^{\prime s'}_{\mb \ma}} \ra
\la Q_s (h_\mc , h_\md ) Q_{s'} (h'_\mc , h'_\md ) \ra ,
\label{eq:spinaverage1}
\end{multline}
where we denote two unit cells (and possibly different imaginary times) with and without a prime, and we introduced
\begin{subequations}
\begin{align}
Q_+ (h_\ma ,h_\mb ) &= 
1 -  C_4 (h_\ma + h_\mb ) ,
\\
Q_- (h_\ma ,h_\mb ) &= 
C_1 +  C_2 h_\ma h_ \mb -  C_3 (h_\ma + h_\mb ) .
\end{align}
\end{subequations}
For the first factor in Eq.~\eqref{eq:spinaverage1} we have
\begin{align}
\la B_{\mb \ma}^s {B^{\prime s'}_{\mb \ma}} \ra
=   \tilde{g}_{11}^2 
+ s s' \tilde{g}_{22}^2
+s'
 \bigl( 
\tilde{g}_{12}^2 
+ s s'
\tilde{g}_{21}^2 \bigr),
\end{align}
so that (because $\tilde{g}_{12}^2 = \tilde{g}_{21}^2$)
\begin{align}
\la V S_z S'_z V^{\dagger} \ra 
 &= \frac{\tilde{g}_{11}^2+\tilde{g}_{22}^2}{32}
\sum_{s=\pm}
\la Q_s (h_3 , h_4 ) Q_{s} (h'_3 , h'_4 ) \ra
\nonumber \\
&+\frac{\tilde{g}_{11}^2-\tilde{g}_{22}^2}{32}
\sum_{s=\pm}
\la Q_s (h_3 , h_4 ) Q_{-s} (h'_3 , h'_4 ) \ra
\nonumber \\
& +\frac{\tilde{g}_{12}^2}{16}
\sum_{s=\pm} s
\la Q_s (h_3 , h_4 ) Q_{s} (h'_3 , h'_4 ) \ra .
\end{align}
The remaining needed averages can also be computed, for example
\begin{subequations}
\begin{multline}
\la Q_+ (h_3 , h_4 ) Q_{+} (h'_3 , h'_4 ) \ra
\\
=
Q^2_+ (\bar{h},\bar{h} ) 
+2 C_4^2 (\tilde{g}_{11} \tilde{g}_{22} - \tilde{g}_{12} \tilde{g}_{21} ),
\end{multline}
\begin{multline}
\la Q_- (h_3 , h_4 ) Q_{-} (h'_3 , h'_4 ) \ra
\\
=
Q^2_- (\bar{h},\bar{h} ) 
+ C_2^2 (\tilde{g}_{11} \tilde{g}_{22} - \tilde{g}_{12} \tilde{g}_{21} )^2
\\
+
2 (C_3 - C_2 \bar{h})^2 (\tilde{g}_{11} \tilde{g}_{22} - \tilde{g}_{12} \tilde{g}_{21} ),
\end{multline}
%
%
%
%\begin{multline}
% \la Q_+ (h_3 , h_4 ) Q_{-} (h'_3 , h'_4 ) \ra 
%% = \la Q_- (h_3 , h_4 ) Q_{+} (h'_3 , h'_4 ) \ra
%=
%Q_+ (\bar{h},\bar{h} ) Q_- (\bar{h},\bar{h} )
%\\
% -2 C_4 (C_2 \bar{h} - C_3)
%(\tilde{g}_{11} \tilde{g}_{22} - \tilde{g}_{12} \tilde{g}_{21} ).
%\end{multline}
\end{subequations}

\subsection{Static spin-spin correlation function}

Let us specialize to the static case at $T=0$ and $i \neq j$, then $\tilde{g}_{11} = - \tilde{g}_{22}$. When $i$ and $j$ are on different (the same) sublattices we have $\tilde{g}_{12}=0$ ($\tilde{g}_{11} = 0$). For different sublattices the result is therefore
\begin{multline}
\la V S_{1z} S'_{1z} V^{\dagger} \ra =
\frac{\tilde{g}^2_{11}}{16} \Bigl( 
Q_+^2 + Q_-^2 
\\
- 2 \bigl[ C_4^2 + (C_3 - C_2 \bar{h})^2 \bigr]  \tilde{g}^2_{11}
+ C_2^2 \tilde{g}^4_{11}
\Bigr) ,
\end{multline}
while on the same sublattice
\begin{multline}
\la V S_{1z} S'_{1z} V^{\dagger} \ra =
\frac{\tilde{g}^2_{12}}{16} \Bigl( 
Q_+^2 - Q_-^2 
\\
+ 2 \bigl[ C_4^2 - (C_3 - C_2 \bar{h})^2 \bigr]  \tilde{g}^2_{12}
- C_2^2 \tilde{g}^4_{12}
\Bigr) .
\end{multline}
For the nearest neighbor spin correlation we may use $\tilde{g}_{11}^2 = \tilde{g}_{22}^2 = - \bar{b}^2$ and $\tilde{g}_{12} = 0$, the result in $d=1$ is plotted in Fig.~\ref{fig:spincorrelations}.
\begin{figure}
\begin{center}
\includegraphics[scale=.6]{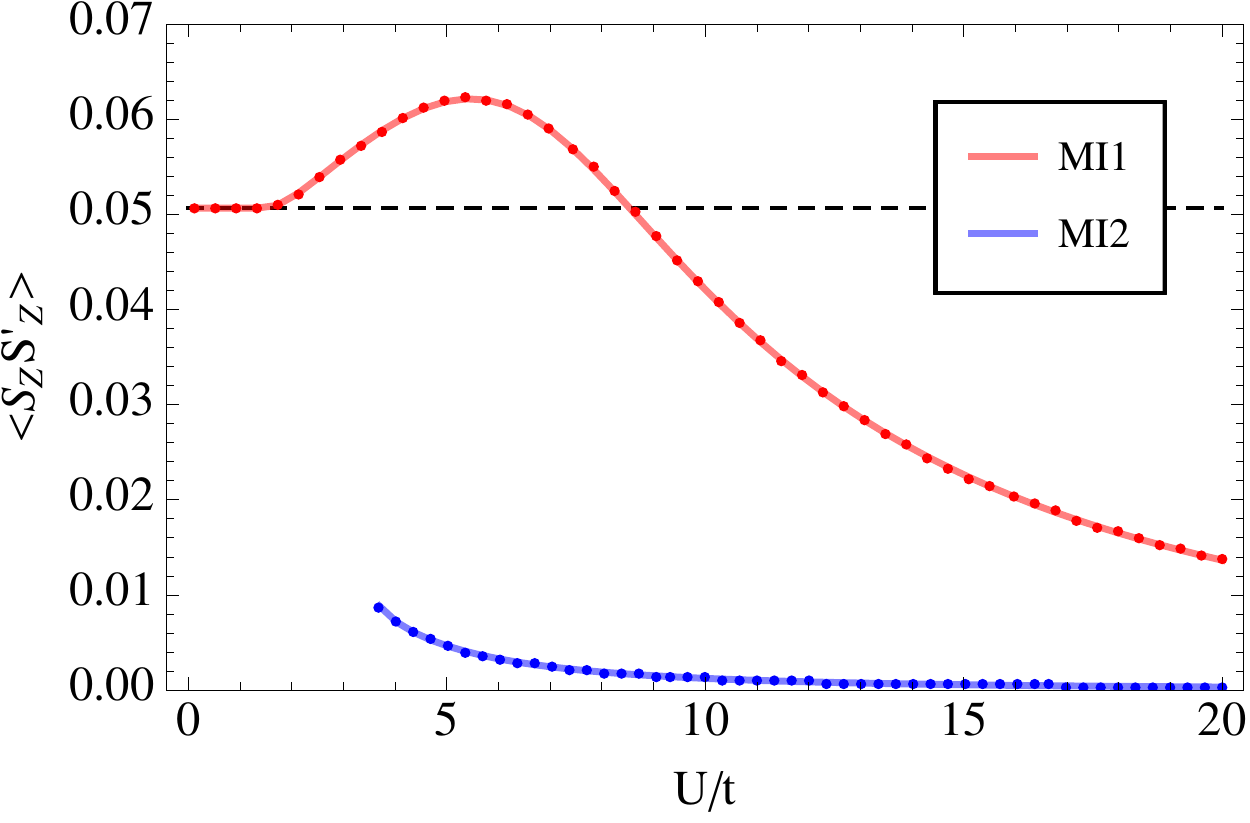}
\end{center}
\caption{Nearest neighbor static spin-spin correlation function $\la S_z S'_z\ra$ in $d=1$ for the two states. The dashed line is the non-interacting result $1/(2\pi^2)$.
\label{fig:spincorrelations}}
\end{figure}

\section{Multi-band Bethe lattice}

On the Bethe lattice in $d$ dimensions the quadratic trial Hamiltonian $\tilde{H}$ can be represented compactly as
\begin{align}
\tilde{H} = -\sum_{\la i,j \ra,\sigma}
\frac{a^{\dagger}_{i\mu\sigma}T_{\mu\nu} a^{}_{j\nu\sigma}}
{\sqrt{d-1}}
- \sum_{i,\sigma} a^{\dagger}_{i\mu\sigma}
\Lambda_{\mu\nu}
a^{}_{i\nu\sigma}.
\label{eq:BethelatticeeffectiveH}
\end{align}
By construction this is also manifestly SO(4)-symmetric when written out in terms of Majoranas when $T$ and $\Lambda$ only connect different sublattices. The matrices $T$ and $\Lambda$ will be determined variationally later. In the following we will drop the diagonal spin index for brevity.

We will now study the effective Hamiltonian in Eq.~\eqref{eq:BethelatticeeffectiveH} using the standard Green's function method \cite{DMFT_RMP1996}. Integrating out everything but two neighboring sites we get an effective $2M \times 2M$ Green function
\begin{align}
G_{2M \times 2M}^{-1} = 
%G_{2s}^{-1} = 
\begin{pmatrix}
\tilde{G}^{-1} & T/\sqrt{d-1} \\
T/\sqrt{d-1} & \tilde{G}^{-1}
\end{pmatrix}
=
\begin{pmatrix}
G_{uc} & G_{nn} \\
G_{nn} & G_{uc} 
\end{pmatrix}^{-1},
\end{align}
where $G_{uc}$ is the Green's function of the unit cell and $G_{nn}$ the Green's function that connects two nearest neighbor unit cells. $\tilde{G}$ is the Green's function of a unit cell when one of the neighbors have been removed. To leading order in $1/\sqrt{d-1}$ this leads to the relations
\begin{equation}
G_{uc} = \tilde{G} , \qquad
G_{nn} = -\frac{1}{\sqrt{d-1}}\tilde{G} T \tilde{G}.
\label{eq:bethelatticegreencomponents}
\end{equation}
A closed equation for $\tilde{G}$ is obtained by integrating out everything but one unit cell, which gives ($z = \io$)
\begin{align}
\tilde{G}^{-1} = z\mathbb{1}  +\Lambda - T \tilde{G} T.
\label{eq:greensfunctionmatrixeq}
\end{align}
One straightforward way to solve this equation, which we use below, is to use the cofactor method in the basis where $T$ is diagonal.

\subsection{Bethe lattice -- 2-band model}

In the 2-band case there are 3 parameters in the most generic time reversal invariant SO(4)-symmetric trial Hamiltonian that we denote $t_1, t_2, \lambda$ so that
\begin{align}
T = 
\begin{pmatrix}
t_1 & 0 \\ 0 & t_2
\end{pmatrix}
,\qquad
\Lambda =
\begin{pmatrix}
0 & \lambda \\ \lambda & 0
\end{pmatrix}.
\label{eq:2bandpmatricesApp}
\end{align}
Using the cofactor method to invert the matrix the solution to Eq.~\eqref{eq:greensfunctionmatrixeq} with the matrices in Eq.~\eqref{eq:2bandpmatricesApp} becomes
\begin{align}
\tilde{G} = \frac{1}{(t_1 t_2 )^2-D^2}
\begin{pmatrix}
z (t_2^2  -D) & \lambda (t_1 t_2+D) \\  \lambda (t_1 t_2+D) & z (t_1^2  -D)
\end{pmatrix},
\end{align}
with $D = 1/\det \tilde{G}$.

Let us first consider the case that $t_2 = 0$ and $t_1 \geq 0$, which is appropriate if one uses mean field theory where $t_2$ vanishes as $d \rightarrow \infty$. In this case the Green's function reduces to
\begin{subequations}
\begin{align}
\tilde{G}(z) &= \frac{1}{D^2}
\begin{pmatrix}
z D & -\lambda D \\ -\lambda D & z (D-t_1^2)
\end{pmatrix},
\label{eq:2bandgreensfunctiont20}
\\
D &=  \frac{z^2-\lambda ^2+\sqrt{\left(z^2-\lambda ^2\right)^2-4 t_1^2 z^2}}{2 }.
\end{align}
\end{subequations}
To have the proper behavior of the diagonal components when $z \rightarrow \infty$ the square root should be picked so that $D \sim z^2$ in this case. The branch cuts are sitting on the real axis between $-z_+ \leq z \leq -z_-$ and between $z_- \leq z \leq z_+$, with
$z_\pm = \sqrt{t_1^2 + \lambda^2} \pm t_1$. This implies that the system is always gapped if $\lambda \neq 0$, this is however not always the case when $t_2 \neq 0$. From the above equations we can extract the weight functions [see Eq.~\eqref{eq:lehmann1}]
\begin{subequations}
\begin{align}
\tilde{A}_{11}(\e) &= \frac{\sqrt{(\e^2-z_-^2)(z_+^2-\e^2)}}{2 \pi t_1^2 |\e|} ,
\\
\tilde{A}_{12}(\e) &= -\frac{\lambda \sqrt{(\e^2-z_-^2)(z_+^2-\e^2)}}{2 \pi t_1^2 |\e|\e} ,
\\
\tilde{A}_{22}(\e) &= 
\frac{\lambda^2 \sqrt{(\e^2-z_-^2)(z_+^2-\e^2)}}{2 \pi t_1^2 |\e^3|} .
\end{align}
\end{subequations}
where the support is $z_- \leq |\e| \leq z_+$.

For reference we also provide some results for the general case that also $t_2$ is nonzero. Let $y = D + (t_1 t_2)^2/D$, for which there is a quadratic equation
\begin{align}
y^2 -(z^2-\lambda^2)y + z^2 (t_1^2+t_2^2) + 2 \lambda^2 t_1 t_2-4 t_1^2 t_2^2 =0 .
\end{align}
The square root should be picked so that $D \sim y \sim z^2$ when $z \rightarrow \infty$; the solution can therefore be represented as
\begin{subequations}
\begin{align}
D &= \frac{y+\sqrt{y^2 - (2 t_1 t_2)^2}}{2} ,
\\
y &= \frac{z^2-\lambda^2}{2}
\nonumber \\
&+\frac{\sqrt{
z^2 (z^2 - 2 \lambda^2 - 4 t_1^2 - 4 t_2^2)
+(4 t_1 t_2 - \lambda^2)^2
}}{2} .
\end{align}
\end{subequations}

\subsection{Nearest neighbor Green's function}

From Eq.~\eqref{eq:bethelatticegreencomponents} we get (again specializing to $t_2 = 0$)
\begin{subequations}
\begin{align}
G_{nn,11} &= \frac{-t_1}{\sqrt{d-1}}\tilde{G}_{11}^2 , \\
G_{nn,12} &=  G_{nn,21}  =  \frac{-t_1}{\sqrt{d-1}}\tilde{G}_{11} \tilde{G}_{12} , \\
G_{nn,22} &=  \frac{-t_1}{\sqrt{d-1}} \tilde{G}_{12}^2 .
\end{align}
\end{subequations}
Using the symmetries we find that this implies
\begin{align}
\la a_{1\vx_i}^\dagger a_{2\vx_i+\vd}^{\,} \ra  = \la a_{2\vx_i}^\dagger a_{1\vx_i+\vd}^{\,} \ra = 0,
\end{align}
independently of the temperature. This is consistent with the imposed particle-hole symmetry.

At zero temperature the frequency integrals needed for the averages can be performed analytically: using formulas 3.152.10, 3.153.7, 3.155.1, 3.156.6 in \cite{Gradshteyn} we obtain
\begin{subequations}
\begin{equation}
\bar{h} = 
\frac{\lambda}{\pi t_1^2} \left[
 \frac{z_+^2 + z_-^2}{z_+} K\Bigl(1 - \frac{z_-^2}{z_+^2}\Bigl)
-2 z_+ E\Bigl(1 - \frac{z_-^2}{z_+^2}\Bigr) 
\right],
\end{equation}
\begin{multline}
\bar{k} = 
\frac{4}{2\pi 3 t_1^3 \sqrt{d-1}} \Bigl[
z_+ (t_1^2 + 2 \lambda^2) E\Bigl(1 - \frac{z_-^2}{z_+^2}\Bigr) 
\\
- \frac{\lambda^2 (3 t_1^2 + 2 \lambda^2)}{z_+}
K\Bigl(1 - \frac{z_-^2}{z_+^2}\Bigl)
\Bigr].
\end{multline}
\end{subequations}
Here $K$ and $E$ are complete elliptic integrals of the first and second kind. Alternative and equivalent formulas can be obtained much more directly from Eq.~\eqref{eq:averageintegrals1} with the Bethe lattice DOS in Eq.~\eqref{eq:dosBethe} after a proper rescaling of the parameters. This gives (recall that $r = t_{12}/\lambda$)
\begin{subequations}
\begin{align}
\bar{h}
 &= 
 \frac{4 \bigl[ (1+r^2) K\bigl(- r^{2}\bigr)
 -  E(- r^{2}\bigr) \bigr]}{\pi r^2} ,
%\end{equation}
%
%\begin{equation}
%\bar{k}
% = \frac{4}{3 \pi r^3 \sqrt{d-1}}
%\Bigl[
% (2 + r^2) E(- r^{2}\bigr)- 2 (1 + r^2) K\bigl(- r^{2}\bigr)
%\Bigr].
%\end{equation}
%\begin{equation}
\\
\bar{k}
 &= \frac{4\bigl[
 (2 + r^2) E(- r^{2}\bigr)- 2 (1 + r^2) K\bigl(- r^{2}\bigr)
\bigr]}{3 \pi r^3 \sqrt{d-1}}
.
\end{align}

\end{subequations}

\section{Transformation classification}

A physical way of understanding and classifying the transformations is to consider, given the imposed symmetries, which states that are allowed to mix. Using charge conservation we know that the total charge $N$ is a good quantum number. Using spin conservation the states can be classified further according to total spin $S_{tot}$ and the spin projection $S_z$. If in one such class with given values of $N, S_{tot}, S_z$ there are $n$ states a generic transformation within that class, which is connected to the identity transformation, can be parametrized with an SU($n$) matrix. The transformation matrices for each spin projection $S_z$ must be the same because of spin rotation symmetry. Demanding that time-reversal invariance is also preserved each matrix must be real and therefore SU($n$) is restricted down to SO($n$). The results of the application of this scheme to local 2- and 3-band models are presented in Tables~\ref{tab:2bandclassification1} and \ref{tab:3bandclassification1}.
\begin{center}
\begin{table}[h]
\begin{tabular}{ |l|l| l|  l|  l|}
\hline
$N$ & $S_{tot}$ & \# / $S_z$ & $N_{c}$ &Transformation \\ \hline
\hline
0 & 0 & 0 & 1  & 1 \\ \hline
1 & 1/2 & 2 & 4 & SO(2) \\ \hline
2 & 0 & 3  & 3 & SO(3) \\
2 & 1 & 1 & 3 & 1 \\ \hline
3 & 1/2 & 2 & 4 & SO(2) \\ \hline
4 & 0 & 0 & 1  & 1 \\ \hline
\end{tabular}
\caption{Classification of states in the 2-band model using total charge $N$, total spin $S_{tot}$, and spin projection $S_z$. Transformations are allowed to mix states with a given value of the spin projection $S_z$ of which there are $\# /S_z$. We also list total number of states in each class $N_c$}
\label{tab:2bandclassification1}
\end{table}
\end{center}
\begin{center}
\begin{table}[h]
\begin{tabular}{ |l|l| l|  l|  l|}
\hline
$N$ & $S_{tot}$ & \# / $S_z$ & $N_{c}$ &Transformation \\ \hline
\hline
0 & 0 & 0 & 1  & 1 \\ \hline
1 & 1/2 & 3 & 6 & SO(3) \\ \hline
2 & 0 & 6  & 6 & SO(6) \\
2 & 1 & 3 & 9 & SO(3) \\ \hline
3 & 1/2 & 8 & 16 & SO(8) \\ 
3 & 3/2 & 1 & 4 & 1\\ \hline
4 & 0 & 6 & 6 & SO(6) \\
4 & 1 & 3 & 9 & SO(3) \\  \hline
5 & 1/2 & 3 & 6 & SO(3)\\ \hline
6 & 0 & 0 & 1 & 1 \\ \hline
\end{tabular}
\caption{Classification of states in the 3-band model using charge and spin. Same convention as in Table~\ref{tab:2bandclassification1}.}
\label{tab:3bandclassification1}
\end{table}
\end{center}

Additional constraints are obtained if we demand that the transformation respects charge conjugation (particle-hole) symmetry. In an $M$-band system this means that the transformations in the charge sectors with $N=n$ and $N = 2M-n$ are related. In particular, if $n \neq 2M-n$ the transformation with $N= 2M -n$ is uniquely determined by the transformation with $N = n$. In the half-filled sector where $N = M$ charge conjugation also splits up the original transformation in an non-trivial way.

In our case the symmetry is even larger and the SO(4) symmetry can be decomposed in terms of spin and pseudospin into SU(2)$\times$SU(2). Each state is then a member of a multiplet in both the spin and pseudospin sectors. The pseudospin symmetry enlarges the discrete charge conjugation symmetry into a continuous SU(2). This breaks down the symmetry further according to Tables~\ref{tab:2bandclassification2} and \ref{tab:3bandclassification2}.

Let us now count the number of transformations in the SO(4)- and time reversal-symmetric case.
According to Table~\ref{tab:2bandclassification2}  a generic transformation in the 2-band model is parametrized by two real numbers. This agrees with the Majorana formulation in the main text where there are two allowed generators.
Consulting Table~\ref{tab:3bandclassification2} we see that the total number of symmetry generators in the 3-band case are those of 2 SO(3) and 2 SO(5) which is
\begin{align}
N_{gc} = 2 \times 3 +2 \times 10  = 26 .
\end{align}
In the Majorana language one can check, most easily using computer algebra \cite{Cliffordpackagearxiv}, that there are 1 bilinear, 8 quadrilinears, and 8 hexalinears. By symmetry there are also 8 octalinears and 1 decalinear. In total there is therefore
 \begin{align}
N_{gm} = 2 \times 1 + 2 \times 8 + 8 = 26,
\end{align}
independent generators. It is reassuring that the two methods give the same number of independent allowed transformations.
\begin{center}
\begin{table}
\begin{tabular}{ |l |l |l |l |l |l |l |}
\hline
$N$ & $S_{tot}$ & \# / $S_z$ & 
$I_{tot}$ & $I_z$ & $N_{c}$ &Transformation \\ \hline
\hline
0 & 0 & 1 & 1 & 1 & 1  & 1 \\ \hline
1 & 1/2 & 2 & 1/2 & 1/2 & 4 & SO(2) \\ \hline
2 & 0 & 2  &  0 & 0 & 2 & SO(2) \\
2 & 0 &  1 &  1 & 0 & 1 & 1 \\
2 & 1 & 3 & 0 & 0 & 3 & 1 \\ \hline
3 & 1/2 & 2 & 1/2 & -1/2 & 4 & inherited from $I_z = 1/2$ \\ \hline
4 & 0 & 1 & 1 & -1 & 1  & 1 \\ \hline
\end{tabular}
\caption{Classification of states in the 2-band model, extending Table~\ref{tab:2bandclassification1} to include the total pseudospin $I_{tot}$ and pseudospin projection $I_z$. The transformation matrix in the same pseudospin multiplet with different $I_z$ must be the same.}
\label{tab:2bandclassification2}
\end{table}
\end{center}
\begin{center}
\begin{table}[h]
\begin{tabular}{ |l |l |l |l |l |l |l |}
\hline
$N$ & $S_{tot}$ & \# / $S_z$ & 
$I_{tot}$ & $I_z$ & $N_{class}$ &Transformation \\ \hline
\hline
0 & 0 & 1 & 3/2 & 3/2 & 1  & 1 \\ \hline
1 & 1/2 & 3 & 1 & 1 & 6 & SO(3) \\ \hline
2 & 0 & 5  &  1/2 & 1/2 & 5 & SO(5) \\
2 & 0 & 1  &  3/2 & 1/2 & 1 & 1 \\
2 & 1 & 3 & 1/2 & 1/2 & 9 & SO(3) \\ \hline
3 & 1/2 & 5 & 0 & 0 & 10 & SO(5) \\ 
3 & 1/2 & 3 & 1 & 0 & 6 & inherited from $I_{z}=1$ \\ 
3 & 3/2 & 1 & 0 & 0 & 4 & 1\\ \hline
4 & 0 & 5  &  1/2 & -1/2 & 5 & inherited from $I_{z}=1/2$  \\
4 & 0 & 1  &  3/2 & -1/2 & 1 & 1 \\
4 & 1 & 3 & 1/2 & -1/2 & 9 & inherited from $I_{z}=1/2$ \\  \hline 
5 & 1/2 & 3 & 1 & -1 & 6 & inherited from $I_{z}=1$ \\ \hline
6 & 0 & 1 & 3/2 & -3/2 & 1  & 1 \\ \hline
\end{tabular}
\caption{Classification of states in the 3-band model using charge, spin, and pseudospin. Same convention as in Table~\ref{tab:2bandclassification2}.}
\label{tab:3bandclassification2}
\end{table}
\end{center}

\end{document}